# Folding Molecular Dynamics Simulations of the Transmembrane Peptides of Influenza A, B M2, and MERS-, SARS-CoV E Viral Proteins

Antonios Kolocouris,[†,*] Isaiah Arkin,[#] & Nicholas M. Glykos[±,*]

[±] *Laboratory of Medicinal Chemistry, Section of Pharmaceutical Chemistry, Department of Pharmacy, National and Kapodistrian University of Athens, Panepistimiopolis, 15771, Greece,  ankol@pharm.uoa.gr*

[#] *Department of Biological Chemistry, The Alexander Silberman Institute of Life Sciences, The Hebrew University of Jerusalem, Edmond J. Safra Campus Givat-Ram, Jerusalem 91904, Israel*

[†] *Department of Molecular Biology and Genetics, Democritus University of Thrace, University Campus, Alexandroupolis, 68100, Greece, glykos@mbg.duth.gr*

# Abstract

Viroporins are small viral proteins that oligomerize in the membrane of host cells and induce the formation of hydrophilic pores in these membranes, thus altering the physiological properties of the host cells. Due to their significance for viral pathogenicity, they have become targets for pharmaceutical intervention, especially through compounds that block their pore-forming activity. Here we add to the growing literature concerning the structure and function of viroporins by studying and comparing —through molecular dynamics simulations— the folding of the transmembrane domain peptides of viroporins derived from four viruses : influenza A, influenza B, and the coronaviruses MERS-Cov-2 and SARS-CoV-2. Through a total of more than 50 μs of simulation time in explicit solvent (TFE) and with full electrostatics, we characterize the folding behavior, helical stability and helical propensity of these transmembrane peptides in their monomeric state and we identify common motifs that may reflect their quaternary organization and/or biological function. We show that the two influenza-derived peptides are significantly different in peptide sequence and secondary structure from the two coronavirus-derived peptides, and that they are organized in two structurally distinct parts : a significantly more stable N-terminal half, and a fast converting C-terminal half that continuously folds and unfolds between α-helical structures and non-canonical structures which are mostly turns. In contrast, the two coronavirus-derived transmembrane peptides are much more stable and fast helix formers when compared with the influenza ones. Especially the SARS-derived peptide appears to be the most fast and stable helix-former of all four peptides studied, with a helical structure that persists almost without disruption for the whole of its 10 μs simulation. We discuss possible interpretations of these findings and their putative connection to the structural characteristics of the respective viroporins.

# 1. Introduction

Over the past decades, an increasing number of both cation- and anion-conducting viroporins have been identified and proposed to play central roles in the viral life cycle, in addition to having a huge impact on pathologies in the host. [1,2] Viroporins are homo-oligomeric proteins with ion channel pores that are formed by α-helical transmembrane (TM) domains. [3] They have been identified in a vast number of pathogenic viruses. Examples are the homotetrameric influenza A (AM2) [4,5] and influenza B M2 (BM2) [6] proton channels and the channels of the severe acute respiratory syndrome coronavirus 2 (SARS-CoV-1, -2), [7] the cause of the ongoing pandemic of COVID-19, and Middle East respiratory syndrome coronavirus (MERS-CoV) [7] that conduct cations. [8–10]

The AM2 and BM2 proteins are 97- and 109-residues single-pass membrane proteins, respectively, that form homotetramers in membrane. [5,11,12] AM2 [13] and BM2 [6] are proton channels, with M2TM being the ion channel pore, [14] and form an active, open state at low pH during endocytosis. In particular, the activation of M2 tetrameric bundle ultimately leads to the unpacking of the influenza viral genome and to pathogenesis. [15] The two proteins share almost no sequence homology except for the HxxxW [16,17] sequence motif in the TM domain that is essential for channel activity with His acting as a sensor residue for proton conduction and Trp as the gate (Figure 2). At acidic pH, the four His residues are protonated, repeal each other and the M2 channel opens and conducts protons. [6,13] Their TM domain sequence arrangements are different, i.e. the AM2TM region encompasses residues 22–46 compared to residues 4–33 in BM2TM (Figure 2). Hence, the unstructured N-terminal segment preceding the TM domain is much longer in AM2. AM2 and BM2 proteins both have relatively large C-terminal cytoplasmic regions. These regions have been suggested to play a role during virus budding [18,19] by recruiting the M2 protein to the cell surface during and viral assembly by contributing to the coating of M2 to the viral envelope. [20,21] The structure of AM2TM (Udorn strain, residues 22-46) homotetrameric bundle has been resolved using both solid state NMR (ssNMR) (PBD ID 2H95 [22], 2L0J [23]) and X-ray crystallography (PDB ID 4QK7 [24]). The structure of influenza BM2 (residues 1-51) homotetramer has been resolved using ssNMR

(PDB ID 6PVR [25]); this construct contains the TM domain (residues 4-33) with residues 34-43 connecting the TM with cytoplasmic domain of full-length BM2. [26]

One shared and conserved viroporin has been identified in SARS-CoV-1, -2 and MERS-CoV the cation-conducting protein E. [8,9,10] It is highly homologous to the deadly SARS-CoV-1 (also known as SARS-CoV), giving rise to the 'SARS' epidemic in 2002 and to the MERS-CoV giving rise to MERS in 2012. [7] E is a 75-residue viroporin that forms a cation-selective channel region across the ERGIC membrane. [8] It has a TM region of 30 residues (8-38) in SARS-CoV-1, -2 and MERS-CoV (SARS ETM and MERS ETM, respectively) with an identical amino acid sequence for SARS-CoV-1 and SARS-CoV-2 (SARS ETM). [27,28] A structure for SARS ETM (residues 8-38, PDB ID 7K3G [28]) has been recently suggested using ssNMR while previous structures were obtained with solution NMR in micelles (residues 8-65, PDB ID 5X29). [27]

The N-terminus of the CoV channels, contains a $(E/D/R)_8 X(G/A/V)_{10} XXhh(N/Q)_{15}$ motif (Figure 2), where h is a hydrophobic residue, and comprises the cation selectivity filter. The protonation equilibria of Glu8, together with the anionic lipids in the ERGIC membrane, may regulate the ion selectivity of ETM at the channel entrance. The third residue of the motif (G/A/V) is conserved among coronaviruses to be small and flexible, which might permit N-terminus motion and/or prevent occlusion of the channel lumen. The last residue of the motif is conserved to be either Asn or Gln, whose polar sidechains can coordinate ions and participate in interhelical hydrogen bonds to stabilize the channel. [29] At the C-terminal part of the TM segment, the conserved small residues Ala32 and Thr35 provide an open cavity for ions. In contrast to these small (or small and polar) residues, the central portion of the TM domain contains four layers of hydrophobic residues, Leu18, Leu21, Val25, Leu28, suggested that narrow the pore radius to ~2 Å. [28] Compared to AM2TM and BM2TM, this much narrower and less hydrated pore can permit only a single file of water molecules, thus partially dehydrating any ions that move through the pore.

It has been shown that in SARS-CoV-1, E mediates the budding and release of progeny viruses and activates the host inflammasome. [30] The expression of SARS-CoV-1 viroporin promotes virus replication and virulence [31] and deletion of the Protein E gene attenuates the virus, resulting in faster recovery and improved survival in infected mice. [10]

Thus, the ion channel activity of proteins AM2, BM2 and E represents a determinant for influenza S, B and SARS-CoV-1 virulence, with the later mirroring the pathology associated with the severe cases of SARS-CoV-2 infection.

Most viroporins have been identified as virulence factors that lead to viral attenuation when deleted. This attenuation is attributed only in part to their channel activity, but nevertheless small-molecule channel inhibitors have been explored triggering also research for the structure determination of the related viroprins. E's channel activity is blocked by hexamethylene amiloride (HMA) [32] and amantadine; [33,34] the latter also inhibits the viroporin AM2. [4,35,36] However, the vast majority of channel inhibitors have been developed against the AM2. [37–41] This is not surprising since AM2 is the best characterized viroporin to date with an established biological role in viral pathogenesis, [15,18,19] combining the most extensive structural investigations conducted without [22,24] or with inhibitors acting as pore blockers, [35,36,42,43] and has emerged as a validated drug target. [35–43] For other viroporins, these studies are still in their infancy.

We have previously showed [44] using circular dichroism experiments that AM2TM is a stable α-helix and by 1.1 μs-molecular dynamics (MD) simulations with adaptive tempering that AM2TM monomer is dynamic in nature and the region encompassing residues C-terminal part (17-25) quickly inter-converts between an ensemble of various α-helical structures, and less frequently turns and coils, compared to the one α-helix for $Ala_{25}$. Our results [44] from Density Functional Theory (DFT) calculations showed that this is due in this lipophilic peptide to CH···O interactions forces between amino acid alkyl side chains and main chain carbonyls, which although individually weaker than NH···O hydrogen bonds, can dissociate and associate easily leading to the ensemble of folded structures observed in folding MD simulations. The CH···O interactions forces have a cumulative effect that can't be ignored and may contribute as much as half of the total hydrogen bonding energy, [44] when compared to NH···O, to the stabilization of the α-helix in AM2TM. Similar folding forces should characterize all lipophilic peptides.

In this work we explored using a total of 50 μs of molecular dynamics simulations with adaptive tempering in TFE, a membrane-mimicking solvent, [45,46] the folding of the TM monomer of four important viroporins. The studied TM peptides are the AM2TM (residues 22-46) and BM2TM (residues 1-31) from influenza, and the SARS ETM and MERS ETM

(residues 8-38) from the two respective coronaviruses. Compared to our previous study with AM2TM using 1.1 μs-MD simulation [44] we applied here longer simulation times to allow for a more ergodic investigation of the conformational space of the peptides. The membrane-mimicking environment [45,46] does limit the amount of interpretation that can be based on these folding simulations and makes the connection with the peptides' behavior in membranes somewhat qualitative in nature.

# 2. Methods

## 2.1 System setup and MD simulation protocol

The preparation of the systems including the starting peptide structures in the fully extended state together with their solvation and ionization states were performed with the program LEAP from the AMBER tools distribution as previously described in detail. [44,47] We followed the dynamics of the peptides' folding simulations using the program NAMD. [48] For all MD simulations we have used periodic boundary conditions with a cubic unit cell sufficiently large to guarantee a minimum separation between the symmetry-related images of the peptides of at least 16Å. We applied the TIP3P water model, [49] the TFE parameterization [50] from the R.E.D. library [51] and the AMBER99SB-STAR-ILDN force field [52–54] which has repeatedly been shown to correctly fold [55] numerous peptides [56–65] including peptides in mixed organic (TFE/water) solvents. [66]

For all MD simulations, adaptive tempering [67] was applied as implemented in the program NAMD. [48] Adaptive tempering is formally equivalent to a single-copy replica exchange folding simulation with a continuous temperature range. For our simulations this temperature range was 280 K to 380 K inclusive and was applied to the system through the Langevin thermostat, as described below. The MD simulations protocol has also been previously described [44,63–65] and in summary was the following. The systems were first energy minimized for 1000 conjugate gradient steps followed by a slow heating-up phase to a temperature of 320 K (with a temperature step of 20 K) over a period of 32 ps. Subsequently, the systems were equilibrated for 1000 ps under NpT conditions without any restraints, until the volume equilibrated. This was followed by the production NpT runs with the temperature and pressure controlled using the Nosè-Hoover [68] Langevin dynamics [69] and Langevin piston barostat [70] control methods as implemented by the NAMD program, [48] with adaptive tempering applied through the Langevin thermostat, while the pressure was maintained at 1 atm. The Langevin damping coefficient was set to 1 ps$^{-1}$, and the piston's oscillation period to 200 fs, with a decay time of 100 fs. The production runs were performed with the impulse Verlet-I multiple timestep integration algorithm as implemented by NAMD, and lasted 10 μs for each of the CoV peptides, and approximately 15 μs for each of the the peptides AM2TM and BM2TM, giving a grant total for the four peptides of 50 μs of simulation time. [48] The inner timestep was 2.5 fs, with short-range non-bonded interactions being calculated every one step, and long-range electrostatics interactions every two timesteps using the particle mesh Ewald method [71] with a grid spacing of approximately 1 Å and a tolerance of $10^{-6}$. A cutoff for the van der Waals interactions was applied at 9 Å through a switching function, and SHAKE [72] (with a tolerance of $10^{-8}$) was used to restrain all bonds involving hydrogen atoms. Trajectories were obtained by saving the atomic coordinates of the whole systems every 1.0 ps.

## 2.2 Trajectory analysis

The analysis of the trajectories was performed as previously described. [44,63–65] Secondary structure assignments were calculated with the program STRIDE. [73] All molecular graphics work and figure preparation were performed with the programs VMD, [74] RASTER3D, [75] PyMol, [76] WebLogo [77] and CARMA. [78]

## 2.3 Statistical significance and sufficient sampling

Folding molecular dynamics simulations, especially when performed with an adaptive tempering protocol, are at the mercy of the enormously complex configurational space encompassed by the unfolded state. The implication is that it is essential to quantify the statistical significance and the extent of sampling of the corresponding trajectories before any conclusions can be drawn from them. In this communication we have quantified statistical significance through a recently described probabilistic method which is based on the application of Good-Turing statistics to estimate how probable it is to observe completely new/unrelated structures if a given simulation were to be extended to longer timescales. [89] The form of the results obtained from this method are shown in Figure 1 for the four trajectories studied in this communication.

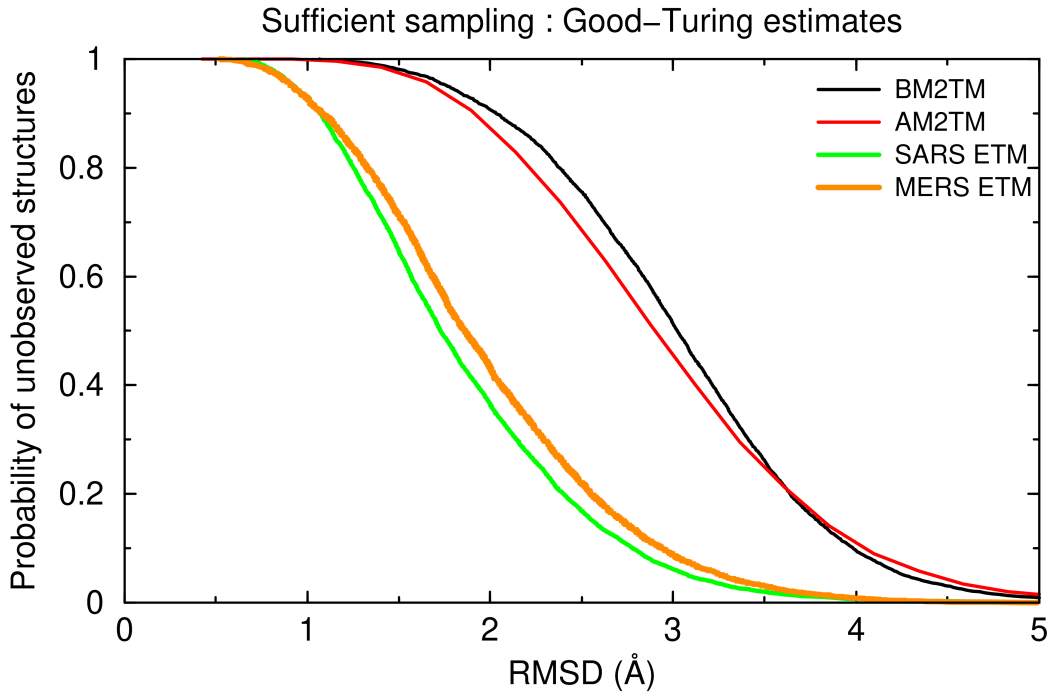

**Figure 1.** Good-Turing estimates for the probability of unobserved structures as a function of the expected RMSD of these structures from the already known (ie. observed in the trajectories) peptide structures. See text for a detailed discussion of this figure.

The information about convergence and extent of sampling is contained in these "Probability vs. RMSD" diagrams which show how probable it is to observe a new structure that would differ by more than a given RMSD from all peptide structures already observed in a given trajectory. All curves start with very high probabilities for low RMSD values, indicating that it is very probable to observe structures that differ only slightly from those already observed. The curves asymptotically approach zero for higher RMSD values, and it is the exact form and how quickly they reach low probability values that informs us just how much structural variability we have *not* yet observed in our trajectories. For the case examined here, the four peptides are clearly clustered in two sets. The coronavirus-derived peptides (green and orange curves in Figure 1), fall-off quite quickly to very low probabilities for RMSD values of around 4 Å. What this implies, then, is that if we were to continue the simulation, we would expect almost all new (previously unrecorded) structures to differ by less than ~4 Å from those already observed. The behavior of the influenza-derived peptides (black and red curves in Figure 1) is significantly different : the curves fall-off slower and maintain significant probability values out to ~5 Å, clearly indicating that a significant volume of the peptides' configurational space has not yet been sampled in these simulations. Note that it is exactly for this reason that we extended the trajectories for the influenza-derived peptides to 15 μs each (instead of 10 μs for the coronavirus peptides).

In summary, the application of Good-Turing statistics allowed us to quantify the extent of sampling in our trajectories and to differentiate between the two sets of peptides based on the structural uncertainty still remaining. The results clearly indicate that with such large uncertainties it would be meaningless to even try to quantify differences between the peptides at the atomic level. A lower resolution comparison, for example at the level of secondary structure stabilities and preferences, it possibly the best that can be achieved with the data available.

# 3. Results

## 3.1 Preliminaries : sequence similarity analysis

Figure 2 shows the amino acid sequences and the corresponding sequence alignment of the four peptides (AM2TM, BM2TM, SARS ETM and MERS ETM) studied here, highlighting their similarities and differences at the sequence level.

There are six residues which are pairwise identical between influenza B and A M2TM peptides (three additional residues are similar ie I, L or V in positions 10, 15, 25 in the BM2TM numbering scheme, Fig. 2). The two influenza A, B peptides have common a HxxxW sequence motif that is considered to include the proton filter and primary gate of the channels. [16,17] The two influenza peptides for the C-terminal half (residues 17-28) share a sequence identity of 33% which is reduced to 15% when the N-terminal half (4-16) is examined. The overall sequence identity between the two peptides is 24%.

The two CoV-derived peptides in their C-terminal half (residues 19-32) share a sequence identity of almost 62%, which drops to 24% for the N-terminal half (residues 2-18). Three additional residues are similar ie I or L in positions 12, 15, 19.

Finally, we note that there are four residues which are pairwise identical between influenza B and either of the CoV peptides but differ when only the two CoV-derived peptides are compared (these four residues are F13, L15, F20 and T24 in the BM2TM numbering scheme, Fig. 2).

```
                      1                                      33
          BM2TM   MFEPFQILSI CSFILSALHF MAWTIGHLNQ IKR
          AM2TM       SSDPLVV AASIIGILHL ILWILDRL
           SARS    ETGTLIVNS VLLFLAFVVF LLVTLAILTA LR
           MERS    RIGLFIVNF FIFTVVCAIT LLVCMAFLTA TR
      Consensus    ......il.. ..fil..lhf .lwtl..L.. ...
```

**Figure 2.** Peptide sequences and alignment for AM2TM (Udorn strain, residues 22-46), BM2TM (residues 1-33), SARS ETM (residues 8-38) and MERS ETM (residues 8-38). Identities and highly similar matches (indicated by a dollar sign in the consensus sequence) are shown in red, pairwise identities in blue.

## 3.2 The simulations indicate the presence of significant differences between the Influenza- and Coronavirus-derived peptides

Figure 3 shows the per-residue secondary structure assignment versus simulation time for each of the four peptides studied. Even a cursory examination of this figure clearly shows that there are pronounced differences between the helical propensity and stability of the four structures.

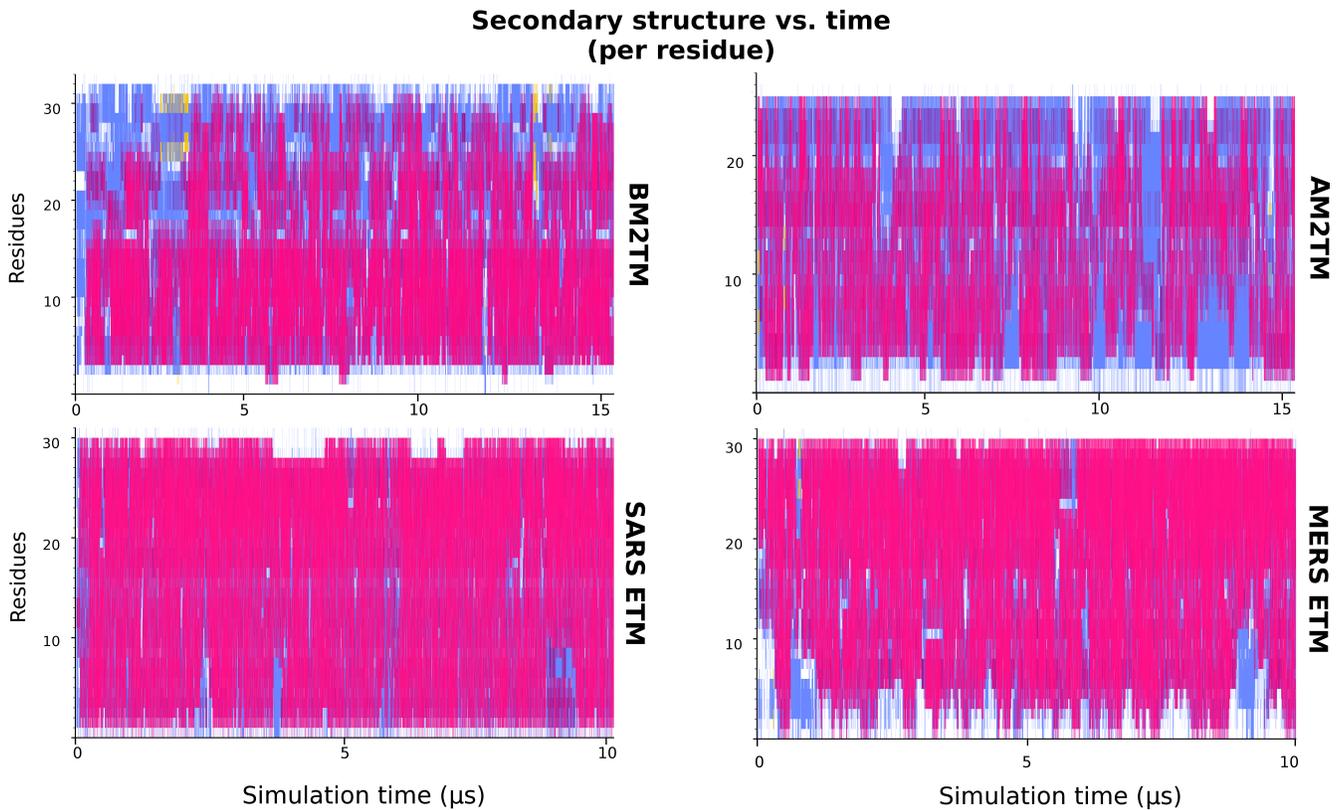

**Figure 3.** Evolution of the per residue secondary structure vs simulation time. The graphs depict the variation of the per-residue STRIDE-derived secondary structure assignments as a function of simulation time for the four peptides. The color coding is red/magenta → α/$3_{10}$ helical structure, cyan → turns, white → coil, yellow → β structure.

Although all four peptides do fold to a mostly α-helical structure as expected, it is the SARS peptide that appears to form an exceedingly stable α-helix (noting here that these results have been obtained from adaptive tempering simulations with the temperature ranging from 280K to 380K). The influenza- and MERS-derived peptides on the other hand, show significant variability –both in helical propensity and helical stability– along the length of their sequences. The influenza-derived peptides appear to show a bipartite organization, with a more stably helical N-terminal half, and a less stable and fast interconverting C-terminal half. This bipartite organization is especially noticeable in the case of the BM2TM peptide. The MERS-derived peptide on the other hand shows the opposite pattern, with a mostly stable and well-behaving C-terminal region, and a more variable N-terminal part.

The least stably folded of all four peptides is the influenza peptide AM2TM which demonstrates significant variability –both in helical propensity and helical stability–along the length of its sequence. This motif of reduced stability of the AM2TM peptide may be connected with the presence of a glycine residue at position 13 (corresponding to the G34 in the 98-residues full-M2 protein) since Gly is known to be a helix-breaker (it has the lowest helix propensity after proline).

To further quantify these observations, we have calculated the fractional helicity for each residue of each peptide over the whole length of their respective simulations. The results from this calculation are shown in Fig. 4. This figure not only places the previous observations on a solid ground, but also highlights silent features that could have been missed from Fig. 3, such as the dip in fractional helicity centered at residue 16 of both CoV-derived peptides. In the paragraphs that follow, we discuss and expand on the different folding behavior of the four peptides.

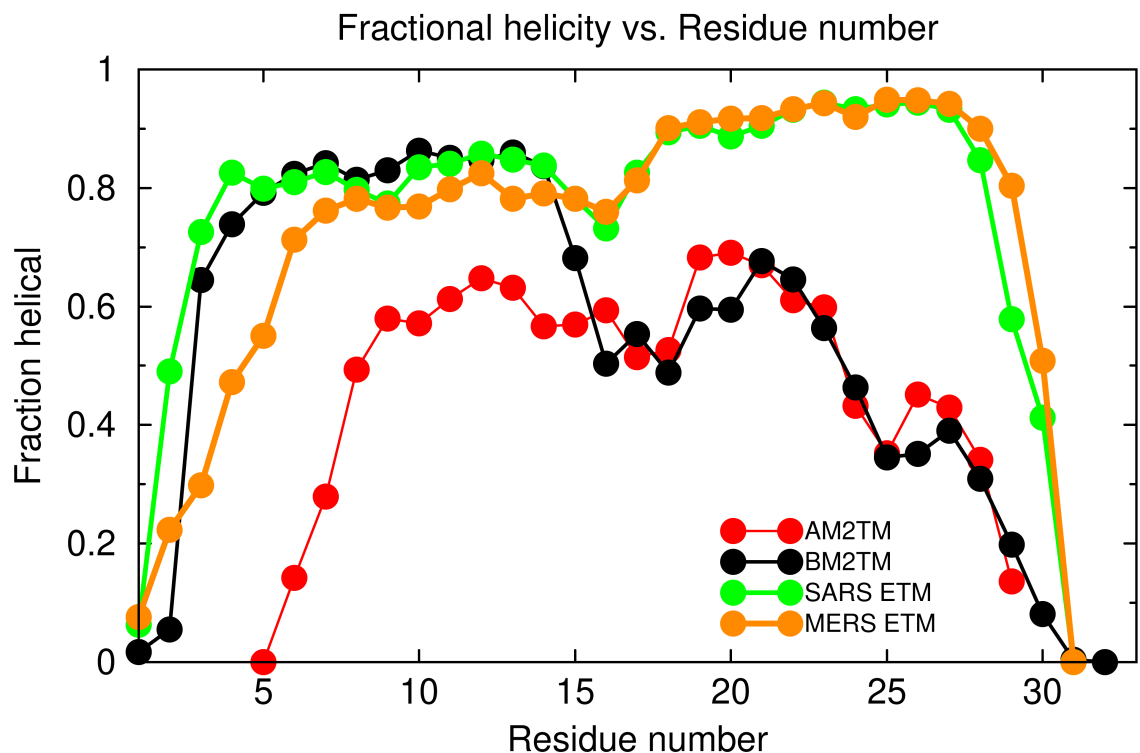

**Figure 4.** Fractional helicity versus residue. The four graphs depict the per residue fractional helicity over the whole length of the three simulations. The color coding is indicated in the figure legend: orange (MERS ETM), green (SARS ETM), black (BM2TM), red (AM2TM). The graphs for BM2TM (black) and AM2TM (red) have been translated by one residue to the left to reproduce the sequence alignment shown in Fig. 2.

## 3.3 The SARS ETM peptide forms an exceptionally stable helical structure

Of the four peptides studied here, the SARS-CoV-derived peptide appears to be the most fast and stable helix-former. Within only ~250 ns of MD simulation time, an almost complete α-helix was formed (noting also that these these are folding simulations which were started from the unfolded/extended state and the results have been obtained from adaptive tempering simulations [67] with the temperature ranging from 280K to 380). This helical structure persists almost without disruption for the whole 10 μs of the MD simulation. There are some helix-fraying events of the termini (see for example the N-terminal fraying events centered at ~2.5 μs, 3.8 μs & 9 μs in Fig. 3), but these do not change the major finding: the SARS-derived peptide is the strongest helix former of the four peptides studied here.

Having said that, there are some salient features of the behavior of the peptide that could have been missed by the data shown in Fig. 3, but are brought forward by the helicity graphs of Figure 4. Referring to this figure, notice the small but systematic difference in helicity between the N-terminal half of the peptide (with a helical fraction of ~0.8) and its C-terminal part (with a helical fraction of ~0.9). As mentioned before, these two parts are separated by a pronounced dip of helicity centered on residue 16. The same pattern is observed for the MERS-derived peptide. The observed pattern for SARS ETM and MERS ETM of two high-helicity parts separated by a dip in helical content is also in agreement with the sequence alignment of the two peptides shown in Fig. 2. There are two regions of significant conservation at the sequence level. The first (N-terminal) region encompasses residues 1-14 (ETGTLIVNSVLLFL in the SARS sequence), followed by four variable residues (15-18, AFVV), and then a second (C-terminal) half which again shows significant sequence similarity between SARS ETM and MERS ETM (19-31, FLLVTLAILTALR in the SARS sequence).

## 3.4 The MERS ETM has a more flexible N-terminal region

As both Figures 3 & 4 indicate, the two CoV-derived peptides are quite similar in their folding characteristics and structural behavior, as would be expected from two peptides sharing significant sequence similarity. Of the four peptides studied here, the next stronger helix former after the SARS-derived peptide is MERS-derived peptide. Thus, the same bipartite organization in two (N- and C-terminal) halves each with a high helical content and separated by a region of reduced helicity near the middle of the peptide is also observed in the MERS-derived peptide.

The similarity is more pronounced in the second (C-terminal) part of the peptides and can be easily identified in Fig, 3 which shows that the helicity of the two CoV-derived peptides is virtually identical in their second half. In the N-terminal part, however, there are significant differences. This finding is not surprising: the peptide sequences at the C-terminal region (residues 19-32) share a sequence identity of almost 62%, whereas in the N-terminal half (residues 2-18) the identity drops to ~24%. As can be seen in Fig. 3 (and in Fig. 4), the first half of the MERS-derived peptide is highly flexible with residues 1-5 all having average helical content of less than ~50%. Why there is such a pronounced difference for the N-terminal residues is difficult to ascertain, as the sequences themselves are closely related (ETGTLIV vs RIGLFIV for the SARS- and MERS- peptides respectively), with the only consistent difference being the substitution of two hydrophobic residues in MERS ETM (I2 & L4) by two threonines in SARS ETM.

## 3.5 The influenza AM2TM peptide is mostly α-helical with an α-helix glycine disruptor at the middle of the peptide

The variability in the AM2TM helix propensity along the length of the peptide is shown in Figs. 2 and 3. Generally, this peptide is highly flexible and continuously folds and unfolds between α-helical structures and non-canonical structures which are mostly turns. There is a discontinuity of helical content at the middle of the peptide which coincides with the presence of a glycine residue (Gly13), which is known to act as a helix breaker. Although highly flexible along its whole sequence, AM2TM shows a more stable α-helical structure from Leu-5 to Leu-19 while at the C-terminal end of the peptide (W20-L25) the helical fraction is significantly lower. Also the capping residues have non-helical $\varphi, \psi$ dihedral angles, although they form helical ($i, i+4$) hydrogen bonds. [79]

A shorter (1.1 µs) molecular dynamics simulation performed previously, [44] had suggested more pronounced differences between the two halves with C-terminal half (after Gly13, sequence ILHLILWILDRL) being much less helical compared to the N-terminal half (sequence SSDPLVVAASII). The longer 15.5 µs-MD simulation described here clearly indicates that the differences in helicity between the two halves is less dramatic than initially estimated.

## 3.6 The Influenza BM2TM peptide is structurally divided in two distinct parts

The BM2TM peptide is significantly different from the three other peptides. It is clearly organized in two structurally distinct parts. The first (N-terminal) region comprises residues 1-15 in the BM2TM numbering. This first half of the peptide demonstrates a strong helix-forming tendency and folds quickly and stably to an α-helix that persists for almost the whole length of the ~15.5 μs simulation. The fractional helicity of this part is identical (if not somewhat higher) than that of the SARS-derived peptide (Fig. 4).

The second (C-terminal) part of the peptide shows a completely different behavior: it is highly flexible and continuously folds and unfolds to transient helical structures interspersed with intervals where it samples non-canonical structures mainly turns but also random coil structures and to few instances β structures (Fig. 3). This flexibility and variability makes a pronounced difference in the fractional helicity graphs of Fig. 4 : almost all residues of the C-terminal half of the peptide have a helical fraction of less 60%. It should be noted, however, that there is some fine structure present in the helical propensity demonstrated by this C-terminal part. As can be seen from both Figs. 3 & 4, residues 20-24 do show an increased preference for a canonical α-helical structure reaching a helical fraction of ~0.7 for residue 22. We are possibly pushing the limits of interpretation of these simulations, but we should note that this motif "High helicity → Dip → High helicity" has been observed on all four peptides studied.

# 4. Summary and Discussion

We performed a total of 50 μs of molecular dynamics simulations with adaptive tempering to study the folding for the transmembrane peptides of the influenza A, B M2, MERS- and SARS-CoV viroporins. The AM2TM and BM2TM peptides have amino acid sequences that differ significantly, both between them, as well as with the two CoV-derived peptides studied in this communication (Fig. 2). On the other hand, the two CoV-derived peptides share a high sequence similarity (Fig. 2). While all peptides are lipophilic, as expected from TM domains, it is worth noting that the AM2TM and BM2TM peptides —which correspond to tetrameric proton channels— are more polar while the two CoV-derived peptides are the most hydrophobic and correspond to viroporins that mediate the conductance of bigger cations, e.g. $Ca^{2+}$.[8–10]

The 15.5 μs-MD simulation of AM2TM revealed that this peptide is highly flexible and continuously folds and unfolds between α-helical structures and non-canonical structures which are mostly turns. It seems however, that the AM2TM prefers an α-helical structure from Leu-5 to Leu-19 and only near the C-terminus of the peptide (from W20 to L25) the helicity is lowered. In AM2TM there is a glycine that acts as an α-helix breaker at the middle of the peptide. In the X-ray structure of AM2TM (PDB ID 4QK7 [24]) or its ssNMR in membrane (PBD ID 2H95 [22]) the structure of the peptide in the tetrameric bundle is α-helical and there is a kink at G13 in the middle of the TM domain, which allows the N- and C-terminal halves of the TM helix to adopt distinct orientations. This G13 kink has been observed also in the ssNMR (PDB ID 2H95,[22] 2KQT[35]) or X-ray (PDB ID 6BKK, [36] 6US9 [80]) drug-bound AM2 structures. This pattern almost exactly matches the the middle of the membrane. Using ssNMR it has been shown that in the tetrameric bundle of AM2TM the helices are flexible with conformational transitions [81,82] that enable protons transportation through the channel.

The amino acid sequence of BM2 does not resemble that of AM2 tetrameric bundle except for the HxxxW motif, where the proton-selective residue is H19 and the gating residue is W23. BM2 has more polar pore-facing residues [17] whereas AM2 has a more hydrophobic pore. Thus, the aqueous pore of the AM2 ion channel is formed by Val[7], Ala[9], Gly[13], His[16], Trp[20], [24] compared with residues Ser$^9$, Ser$^{12}$, Ser$^{16}$, His$^{19}$, Trp$^{23}$ that line the pore of the four-helix bundle in BM2. [17]

While both BM2 and AM2 channels exhibit microsecond-timescale His and Trp sidechain motions, similar to AM2, [83] the BM2 peptide lacks the alternating-access hinge motion but instead, it opens through a scissor motion. Upon activation, BM2 expands its pore along the entire channel, while AM2 constricts its N-terminus but expands its C-terminus. AM2 converts between two conformations: an N terminus-dilated and C-terminus-constricted (N$_{open}$-C$_{closed}$) conformation that is dominant at high pH [84,85] and an N terminus-constricted and C-terminus dilated (N$_{closed}$-C$_{open}$) conformation that is dominant at low pH. [86,24]

An interesting question here is which amino acid sequence features cause the alternating-access motion and the asymmetric conductance of AM2, and their absence in BM2 at acidic pH where His residues are protonated. [25] In BM2 the G13 is replaced by S16, which reinforces the helical backbone and prevents separate motion of the two halves of the BTM helix. Additionally, BM2 has a symmetric HxxxWxxxH motif that is absent in AM2. Therefore, the electrostatic properties of the C-terminal residues in BM2 together with the absence of a central flexible Gly, likely explain the symmetric backbone scissor motion of BM2 for channel activation and the consequent bidirectional proton conductance. These experimental observations are also consistent with the increased flexibility observed for the C-terminal half of BM2 in our simulations.

The BM2TM simulation showed that the N-terminal half of the peptide (comprising residues 1-15) folds quickly and stably to an α-helix with a helical fraction of ~0.8, which is identical (if not somewhat higher) than the helicity observed in the two CoV-derived peptides. In contrast, the second (C-terminal) part of the peptide has a helical fraction lower than 0.6, is more flexible compared to AM2TM and continuously folds and unfolds to transient α-helical structures, turns and coiled-coil and instantaneously β-structures (noting, however, that residues 20-24 do demonstrate an increased preference for a canonical α-helical structure).

The SARS-derived peptide appears to be the most fast and stable helix-former with a helical structure that persists almost without disruption for the whole 10 μs of the MD-simulation. The N-terminal half of the peptide (with a helical fraction of ~0.8) is separated by the C-terminal part (with a helical fraction of ~0.9). Indeed, the first (N-terminal) region encompasses residues 1-14 (ETGTLIVNSVLLFL, corresponding to 8-21 in the SARS sequence), followed by four variable residues 15-18 (AFVV, corresponding to 22-25 in the SARS sequence), and then a second C-terminal half with residues 19-31 which again shows significant sequence similarity (FLLVTLAILTALR, corresponding to 26-38 in the SARS sequence) with a pronounced dip of helicity centered on residue 16 (F, residue 23 in the SARS sequence).

Compared with SARS-CoV-2 ETM, the MERS ETM peptide has a pronounced identical helicity in the second (C-terminal) half, which is not surprising given that the peptide sequences for the region 19-31 share a sequence identity of almost 60%. In contrast, the N-terminal part of the MERS-derived peptide is highly flexible with residues 1-5 all having average helical content of less than ~0.5 although also the N-terminal sequences are closely related (ETGTLIV vs RIGLFIV for the SARS- and MERS- peptides respectively), with the only consistent difference being the substitution of two hydrophobic residues in MERS ETM (I2 & L4) by two threonines in SARS ETM.

The peptides' secondary structure preferences and dynamics reflect features of the quaternary organization of the corresponding proteins and their biological function. The ETM helical bundle of SARS-CoV-2 is compact and rigid, while AM2 and BM2's TM domains, which have a higher percentage of polar residues such as His and Ser, form wider and more hydrated pores. [35,24,25] Indeed the ETM peptides are more immobilized than M2TM peptides. This immobilization suggests that ETM compared to M2TM pore may interact extensively with lipids. [87,88] Finally, the helix distortion at residues Phe20–Phe23 may cause the two halves of the ETM protein to respond semi-independently to environmental factors such as pH, charge, membrane composition and other viral and host proteins.

At the immobilized C-terminal end of the TM segment, the conserved small residues Ala32 and Thr35 provide an open cavity for ions. In contrast to these small (or small and polar) residues, the central portion of the TM domain contains four layers of hydrophobic residues, Leu18, Leu21, Val25 and Leu28, which narrow the pore radius to ~2 Å. This narrow pore can permit only a single file of water molecules, thus partially dehydrating any ions that move through the pore.